\documentclass[letter]{aa} 
\usepackage{natbib}
\usepackage{graphicx}
\usepackage{txfonts}
\begin{document}

\title{The  young stellar population at the center of NGC~205}

   \author{L. \,Monaco\inst{1,2}, 
   I. \,Saviane\inst{1}, 
   S. \,Perina\inst{3}, 
   M. \,Bellazzini \inst{3},
   A. \,Buzzoni \inst{3},\\
   L. \,Federici \inst{3},
   F. \,Fusi Pecci \inst{3},
         \and
   S. \,Galleti \inst{3}
          }

\institute{
European Southern Observatory, Casilla 19001, Santiago, Chile
\and
Universidad de Concepci\'on,
Casilla 160-C, Concepci\'on, Chile
\and
Istituto Nazionale di Astrofisica --
Osservatorio Astronomico di Bologna, Italy
I-40127 Bologna, Italy
}

\authorrunning{Monaco et al.}
\mail{lmonaco@astro-udec.cl}

\titlerunning{ACS photometry of NGC205}

   \date{Received ; accepted}

 
  \abstract
  {NGC~205 is a peculiar dwarf elliptical galaxy hosting in its center a
   population of young blue stars. Their origin is still matter of debate, the
   central fresh star formation activity possibly being related  to dynamical 
   interactions between NGC~205 and M31.} 
    {The star formation history in the central 30\arcsec ($\sim$120~pc) around
     the NGC~205 central nucleus is investigated in order to obtain clues to the
     origin of the young stellar population.}
   {Deep HST/ACS CCD photometry is compared with theoretical isochrones and
   luminosity functions to characterize the stellar content of the region under
   study and compute the recent SF rate.}
   {Our photometry reveals a previously undetected blue plume of young stars
   clearly  distinguishable down to I$\simeq$26. Our analysis suggests that
   1.9$\times$10$^5$~M$_\odot$ were produced between approximately 62~Myr and
   335~Myr ago in the NGC~205 inner regions, with a latest minor episode
   occurring $\sim$25 Myr ago. This implies a star formation rate of 
   $\sim$7$\times$10$^{-4}$~M$_\odot$/yr over this period.  }
   {The excellent fit of the observed luminosity function of young main 
   sequence stars obtained with a model having a constant star formation rate
   argues against a tidally triggered star formation activity over the last
   $\sim$300~Myr. Rather, a constant SF may be consistent with NGC~205 being on
   its first interaction with M~31.}


   \maketitle
%

\section{Introduction}

NGG~205 is one of the brightest \citep[M$_V$=-16.6][]{mateo98} M~31 satellites
and is a peculiar nucleated dwarf elliptical galaxy. It hosts a fairly complex
stellar content. Color-magnitude diagrams and time series revealed the presence
in NGC~205 of RR~Lyrae and carbon stars, i.e. both old and intermediate age
stars as well as sequences of asymptotic  and red giant branch stars
\citep[AGB/RGB,][]{sharina06,richer84,davidge03,davidge05}. \citet[][]{lee96}
broadly summarized the NGC~205 star formation (SF) history. 

The presence of a population of bright blue stars in the most central region of
NGC~205 has been known since the early studies \citep[][and references
therein]{hodge73} on this galaxy and is an unusual characteristic for a dwarf
elliptical \footnote{On the other hand, young nuclear clusters seem relatively
frequent  in nucleated dwarfs \citep{rossa}.}. \citet[][hereafter
C99]{cappellari99} revealed that many of the brightest  blue sources classified
as very young stars \citep{peletier} were in fact multiple systems, clusters or
star associations. \citet[][hereafter D03]{davidge03}, studying the population
of Asymptotic Giant Branch (AGB) stars, noted that multiple episodes of SF may
have occurred in the most central regions \citep[see also ][]{bica90,lee96} with
a time spacing compatible with the putative NGC~205 orbital period
\citep[][]{cepa88}.  Therefore, tidal interactions with M~31 could have
triggered the latest episodes of SF. Indeed, evidence of past dynamical
interactions of NGC~205 with its bright companion has been accumulating over the
years \citep{mccon04,geha06,der06} and residual gas \citep{welch98} and dust
\citep{marleau06} have also been detected in NGC~205 \citep[see
also][]{davidge05}.

However, in spite of the great interest in the recent star formation history in
the innermost region of this galaxy, previous attempts to directly analyze
the young Main Sequence population have been unsuccessful, likely due to the
extreme degree of crowding affecting the region of interest \citep{butler05}.
Here we use the exceptional spatial resolution of the Advanced Camera for Survey
(ACS) on board the Hubble Space Telescope (HST) to derive accurate  stellar
photometry in the circum-nuclear region of NGC~205, resolving for the first time
the young Main Sequence (MS) population down to I$\simeq 26$. 


\section{Data reduction}

The inner 29$\arcsec\times$26$\arcsec$ around the central nucleus of NGC~205
were imaged using the ACS high resolution channel (HRC) and retrieved from the
archive through the MAST web interface\footnote{http://archive.stsci.edu/}.  At
the distance of NGC~205 \citep[824~kpc, see][]{mccon05}, the observed field of
view corresponds to about 120~pc. The data were obtained during Cycle~11 
(program 9448, PI L. Ferrarese). F555W and F814W frames were taken  for a total
integration time of 2560s and 2440s, respectively.  Images were acquired at four
different pointings to improve the resolution and a pixel scale of
0\farcs022~px$^{-1}$ was eventually obtained \citep[see][for further
details]{acs}.  Photometry was performed using the ACS module of 
Dolphot\footnote{http://purcell.as.arizona.edu/dolphot/}, with the same approach
described in \citet{silvia} and \citet{sibilla}. A final photometric catalog  of
about 26000 stars was built, retaining only objects classified as {\em
bona-fide} stars (i.e., quality flag =1) and having a $\chi^2<$2.5. This
selection allows us to exclude several spurious and/or poorly measured sources
without a significant loss of information \citep{silvia}. Dolphot automatically
transforms instrumental magnitudes into the VEGAMAG and the Johnson-Kron-Cousins
BVRI systems adopting the  calibrations by \citet{sirianni}. In the following we
will always adopt Johnson-Kron-Cousins  V,I magnitudes. 

The majority of the central nucleus is unresolved and stars begin to be measured
at radial distances $\geq 0\farcs9$ (about 40~px) from the cluster center. At a
radius of $\sim$5$\arcsec$, the nucleus surface brightness profile  joins the
profile of the underlying galaxy \citep[see][]{acs,butler05,jones}. There is no
particular difference in the stellar population in the annulus $1\arcsec\la r\la
5\arcsec$, where the resolved outskirts of the nuclear cluster are mixed with
the underlying population, and the surrounding field. Therefore, in the
following, we will consider the whole sample as representative of the
circum-nuclear region of the galaxy.

\subsection{Comparison with previous photometries}

Fig.~\ref{iso} shows the obtained I {\it vs} V-I color-magnitude diagram (CMD).
WFPC2 photometry of the central region of NGC~205 was presented by
\citet[][hereafter BM05, their fields F3 and F4]{butler05}. An inspection of
Fig.~\ref{iso} and their Fig.~4 reveals that the ACS photometry presented here
is significantly deeper than any other obtained so far for the same field. Also,
evolutionary marks like the clump of stars at V-I$\simeq$0.9; I$\simeq$21.8
(hereafter YC, young clump, see Sect.~3) have never been detected before. 

While the quality of the WFPC2 and ACS photometries are comparable along the RGB
(V-I$\geq$1), the populated young MS (blue plume, BP) visible in Fig.~\ref{iso}
at V-I$\simeq$-0.2  down to the detection limit is practically absent in WFPC2
CMDs. Quite likely, such blue stars do not exist at all in the more external
regions of NGC~205. However, they could have been detected in the F3 and F4
fields, which encircle the ACS area. 

Blending of red stars may appear in the photometry as sequences of spurious blue
stars. Extensive artificial star tests have been performed following the
procedure described in \citet[][]{sibilla} to assess the completeness and the
impact of blending on our photometry as a function of the star color and
magnitude. The bin-migration effect is quite limited for stars having V-I$<$0.5.
In the magnitude range 22$<$I$<$24.5, only a fraction between 2\% and 4.5\% of
the injected stars change their magnitude by $\Delta$I$\geq$-0.5 when recovered.
This means that about 64 in the $\sim$1764 stars under consideration
($\sim$3.6\%) may be seriously affected by blending of star pairs. We are thus
fully confident about the genuine nature of the detected young MS and that
blending effects do not jeopardize the results presented. The lack of such BP in
the BM05 photometry is likely to be due to the details of the photometry process
and to the selection criteria adopted to filter the detected sources.

It has long been known that the brightest blue stars in NGC~205 are concentrated
in a region of $\sim$300~pc (100\arcsec) around the nucleus \citep{hodge73}. 
Several authors have presented integrated photometry surface brightness
profiles  over extended areas in NGC~205 \citep[see][and references
therein]{lee96}.  The galaxy colors become bluer inward within $\sim$50\arcsec
from the center with  the exception of the inner $\sim$2\arcsec where the colors
become redder. Most of the bright blue stars were found to be concentrated
within a region  of $\sim$20\arcsec diameter around the central nucleus
\citep[][]{peletier}. However, as already mentioned, many of these stars were in
fact clusters  or star associations, as pointed out by C99 using WFPC2 data.

The spatial distribution of BP stars detected in our ACS photometry, on the
other hand,  does not present any obvious clustering, i.e. they do not belong to
associations  similar to the ones discovered by C99 (which are located just
outside the limits  of our ACS field). The same applies also to stars belonging
to the YC.

\section{Star formation history}

\begin{figure}
  \centering
  \includegraphics[width=9cm]{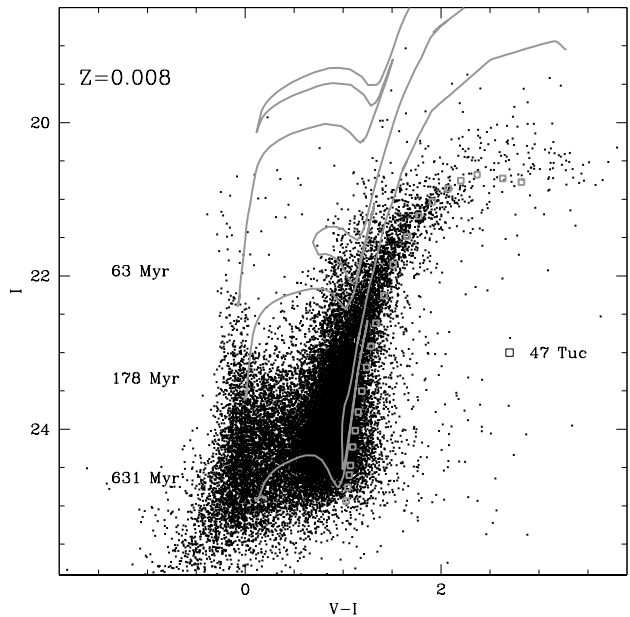}

     \caption{Theoretical post MS isochrones (Z=0.008) from the \citet[][]{leo}
     library superimposed on the ACS photometry.  The mean ridge line of the
     galactic globular cluster 47~Tucanae is also plotted as open squares.}

	\label{iso}
  \end{figure}

An overall idea of the star formation history in the observed NGC~205 regions
can be readily obtained with the aid of theoretical isochrones.  On top of the
ACS photometry, we plotted in Fig.~\ref{iso} a selection of isochrones (in the
post MS phase) from the \citet[][]{leo} library  and the mean ridge line of the
galactic globular cluster 47~Tucanae (open squares) from \citet[][]{ivo00}.
Reddening and distances for this cluster were taken from the most recent version
of the \citet[][]{harris} catalogue.

We adopted for NGC~205 the following parameters: (m-M)$_0$=24.58, E(B-V)=0.08
from \citet[][]{mccon05} and \citet[][]{sfd}, respectively. However, this
reddening value does not account for the internal extinction nor for the
contribution from M31's dust clouds in front of NGC~205. While it has been
suggested that NGC~205 might even lie in front of M31 \citep[][]{howley}, some
degree of differential reddening is certainly present in the galaxy region under
consideration. Nevertheless, its overall effect on the photometry presented here
is negligible (see BM05 for a  discussion).

Fig.~\ref{iso} shows that the blue plume (V-I$\simeq$-0.2) is dominated by MS
stars likely older than $\sim$50-60~Myr and reaching $\sim$650~Myr at the
detection limit. A sparse population of younger stars is also visible at
I$\leq$22. They might correspond to a $\sim$25~Myr old simple stellar population
(see below). The YC may mark instead a star formation episode occurring some
$\sim$200~Myr ago and is made up of blue-loop helium burning stars. The bulk of
the RGB population is matched at its red edge by the ridge line of the old and
relatively metal rich globular cluster 47 Tuc \citep[Fe/H=-0.70,][]{ivo00}, and
is broadly compatible with Z=0.008 isochrones a few Gyr old. 
\citet[][]{sharina06}, using the Lick indexes, derived  a mean metallicity of
[Z/H]$\simeq$-0.5, age of 1.9~Gyr  and no alpha-enhancement for their most
central field outside the nucleus \citep[see also][BM05]{mould84}.  This is also
consistent with the presence of a well populated red-clump along the RGB at
I$>$24 (see also BM05) which, at odds with the BP population, is a dominant
feature also in the CMDs of external regions of NGC~205 (see BM05). 

\subsection{Recent star formation history}\label{recent}

The luminosity function (LF) of the MS can be used to derive the mass in young
stars and the star formation rate (SFR) at late epochs in the central region of
NGC~205 \citep[see][hereafter SHR04]{sfr}. We assume for the moment that the  SF
has been continuous and constant (but see next section) for the last few hundred
Myr. Thus, the overall SF activity can be approximated by the sum of a discrete
number of simple stellar populations (SSPs).  Each generation of stars will
produce a power law LF which is added to that of the previous generations. The
observed LF will result, then, from the convolution of the LFs of the SSPs (see
SHR04). 

\begin{figure}
  \centering
  \includegraphics[width=9cm]{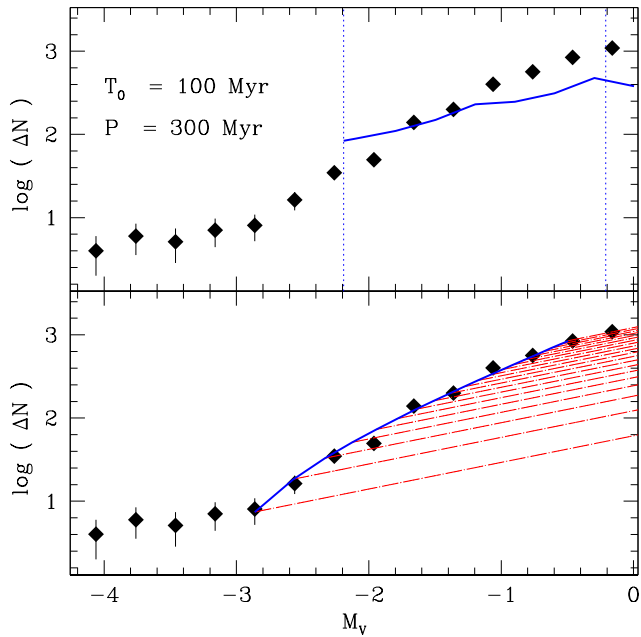}

     \caption{Lower panel: logarithmic integrated LF of young MS stars (filled
 diamonds). The LF is reproduced by the sum (thick continuous line) of 20~LFs 
 (dot-dashed lines) of simple stellar populations.  It is assumed here that the
 slope of the single LFs is equal to that of NGC 2004 (a young populous cluster
 of the Large Magellanic Cloud; $\alpha$=0.32$\pm$0.04, see Fig. 13 in SHR04)
 and that the simple stellar population contains 0.5 times the number of stars
 in NGC~2004. The populations were generated at a constant rate during
 $\sim$273~Myr (see \S\ref{recent}), and all stars now evolved off the main
 sequence were removed from the LFs.  Upper panel: the observed LF (filled
 diamonds) compared with a synthetic one (thick continuous line) generated with
 a bursting star formation. The epoch of the last episode (T$_0$) and the time
 lapse between the star formation episodes (P) are indicated. Dotted vertical
 lines mark the termination of the MS for the single bursts. See SHR04 and text
 for details.} 

	\label{sfr}
  \end{figure}

We select BP stars in the color region at V-I$<$0.2. Besides MS stars, a
negligible fraction of blue loop stars may also be present in the selected
sample. In the color range under consideration, the completeness factor is
$\sim$0.66 at I=24.5 and a comparison of the surface brightness profile
presented by \citet[][]{acs} with a star-count based one reveals that no radial
variations of the completeness factor are present for I$\leq$24.5, as long as
radial distances greater then  r$>5\farcs1$ from the central nucleus are
considered. Therefore, in the following, only stars having  I$\leq$24.5 (i.e
M$_V\leq$-0.5) and r$>5\farcs1$ will be included in the analysis. Fig.~\ref{sfr}
shows the logarithmic cumulated LF of the selected stars (diamonds), corrected
for the appropriate completeness factors.   

In order to estimate the mass in young stars, we followed the same approach as
SHR04 (see lower panel in Fig.~\ref{sfr}).  For stars fainter than
M$_V\simeq$-3.0 (i.e. I$\simeq$22), the observed LF is well reproduced (heavy
solid line) by the sum of 20 LFs of simple stellar populations (dot-dashed
lines). The slope of the single LFs is assumed equal to that of the young Large
Magellanic Cloud cluster NGC~2004 (see SHR04 for details). Each SSP contains
half of the stars in NGC~2004.  The populations were generated at a constant
rate, and all stars now evolved off the main sequence were removed from the LFs.
This way we obtain that  $\sim$1.9$\times$10$^5$~M$_\odot$ were produced between
$\sim$62~Myr and $\sim$335~Myr ago. Therefore, the star formation rate over this
$\sim$273~Myr lapse has been $\sim$7$\times$10$^{-4}$~M$_\odot$/yr. Note,
however, that the star formation should have been active since at least
$\sim$650~Myr ago (see Fig.~\ref{iso}). 

The sparse population of stars brighter than I$\simeq$22 (M$_V<$-3) presents
instead a flatter LF which may be compatible with an isolated SF episode. The
age of this population can be estimated by assuming that the brightest stars are
near the termination of the MS. Using formula B1 of SHR04, their absolute
luminosity (M$_V\simeq$-4.2) yields an age of  $\sim$25~Myr. Hence, this
population would be slightly younger than the  clusters studied by C99 (50~Myr
and 100~Myr, respectively). 

Relying on the \citet[][hereafter B88]{burstein88} IUE data, the UV flux at
1550~\AA\ leads  to a galaxy luminosity of  $\log L_{\rm 1550} =
23.63$~ergs\,s$^{-1}$\,Hz$^{-1}$ across the IUE field of view. Such a luminosity
corresponds to a current SFR~$\lesssim 10^{-5}$~M$_\odot$\,yr$^{-1}$, according
to the \citet[][]{buzzoni02} calibration. The IUE data indicate, therefore, a
drastic reduction of galaxy SF at the current epoch.

Based on a further set of IUE observation, that sampled the nucleus of  NGC~205
(region ``N"), and the region 1\arcmin~north of it (region ``B"),
\citet[][]{iue} estimated a mass of young stars (M$\geq$1~M$_\odot$) of 2 and 
1$\times$10$^5$~M$_\odot$ for the two galaxy regions, respectively. Region B 
corresponds to a part of NGC~205 where some bright blue stars were observed but
does not overlap with our field. Their region N value and ours are similar.
However, their value includes the contribution from the central nucleus which, 
instead, is excluded in our LF-based estimate. Furthermore, they estimate the
mass of all stars having M$\geq$1~M$_\odot$, while we neglect the contribution
by star formation episodes older than 335~Myr, as they do not show up in the LF
in the selected magnitude range (see Fig.~\ref{sfr}). On the other hand, the IUE
field of view is equivalent to a circular aperture with a diameter of 14\arcsec
(see B88), and, as such, it corresponds to about a fifth of the observed ACS
field.

\section{Summary and discussion}

We presented new, deep ACS photometry of the inner 30\arcsec ($\sim$120~pc)
around the NGC~205 nucleus. The most notable feature presented here is the well
populated blue plume of young MS stars visible at V-I$<$0.2 down to the
detection limit (I$\simeq$26, see Fig.~\ref{iso}). Previous photometries were
unable to probe this population, which is confined to the central galaxy
regions. The LF of blue plume stars has been used to investigate the recent star
formation in the central region of NGC~205. We found that
$\sim$1.9$\times$10$^5$~M$_\odot$ were produced between $\sim$62~Myr and
$\sim$335~Myr ago, corresponding to a star formation rate of
$\sim$7$\times$10$^{-4}$~M$_\odot$/yr. However, star formation has been active
since at least $\sim$650~Myr ago, which corresponds to our detection limit (see
Fig.~\ref{iso}). A small number of $\sim$25~Myr old stars are also present, in
agreement with previous findings by C99 and \citet[][]{lee96}.

BM05 and D03 concluded that the latest star formation episodes occurred a few
10$^8$~yr ago in NGC~205. This led them both to speculate that the recent star
formation in NGC~205 might have been triggered by past interactions with M~31.
This is certainly an intriguing possibility.  In fact, \citet[][hereafter
CB88]{cepa88} estimated the orbital period of NGC~205 in $\sim$300~Myr, with the
last passage through the M31 disk occurring $\sim$100~Myr ago.  However, the LF
of the young MS is compatible with a constant SF rate, at least over the last
$\sim$300~Myr (see Fig.~\ref{sfr}, lower panel). 

In Fig.~\ref{sfr} (upper panel), we compare the young NGC~205 MS LF (filled
diamonds) to the LF expected in the presence of a bursting star formation
activity (thick continuous line). The BP mean ridge line and its broadening were
adopted to generate a synthetic MS. For each SF episode, the termination of the
MS was determined  using formula B1 in SHR04. Each burst is assumed to produce
the same number of stars and an NGC~2004-like LF. The resultant - synthetic - LF
is the sum of the stars generated in the single bursts, normalized to the total
number of observed stars. 

We assumed bursts to occur with a time spacing equal to the orbital period
(P=300~Myr), the last SF episode having occurred at the epoch of the last
passage, T$_0$=100~Myr ago. The vertical lines mark the termination of the MS of
the various SF episodes. Given the above parameters, most of the magnitude range
under consideration is covered by just one episode of SF (see Fig.~\ref{iso})
and the synthetic LF is clearly not compatible with the observed one. More in
general, a bursting SF with a period longer than $\sim$100~Myr would imply an LF
flatter than observed. On the other hand, a series of closely spaced star
formation episodes would approximate a continuous star formation (lower panel)
and, as expected, we find that the observed LF is well reproduced by a series of
episodes spaced by only 10-20~Myr, the last one occurring $\sim$63~Myr ago (see
also Fig.~\ref{iso}). This kind of activity is reminiscent of the stochastic
self-propagating star formation  theory proposed by \citet[][]{gs78} and later
discussed, e.g., by \citet[][]{shore83} and \citet[][]{valle05}.

Furthermore, the star formation has certainly been active since more than
$\sim$650~Myr ago (see Fig.~\ref{iso}) and no indication of a significantly
enhanced SF activity at any particular epoch is detected. The only feature that
may be connected to an enhancement of the SF rate over a timescale comparable
with the expected orbital period is the YC, that is presumably associated with a
short episode that occurred $\sim$200 Myr ago.  Therefore, our photometry
does not lend support the hypothesis of a tidally triggered SF at late epochs in
NGC~205. Rather, a continuous SF may be consistent with NGC~205 being in its
first interaction with M~31, as recently proposed by \citet[][]{howley}.

The presence of a young population in a dwarf elliptical (over spatial scales
much larger than the nucleus) remains puzzling, as is the significant amount of
gas observed in this galaxy. The detected amount of gas, however, seems to be
compatible with being returned to the interstellar medium by evolved stars
during a burst of star formation starting $\sim5\times10^{8}$~yr ago
\citep{marleau06}. The present letter shows that the SF was indeed active on a
similar time scale. 

\citet[][]{lisker_etal06} discovered a number of dE with disk-like features in
the Virgo cluster and introduced the term ``dEdi'' for those galaxies. Their
analysis supports the idea that dEdis is a population of genuine disk galaxies
and not just spheroids hosting a disk, perhaps only of tidal origin. These
authors also flag NGC~205 as dEbc type, a subclass of the dEdis having blue
centers \citep[see][]{lisker_etal07}. In fact, the large scale dynamics reveals
that NGC~205 is at least partly supported by rotation \citep{der06,geha06}, with
rotation being detected only along the major axis and \citet[][]{der06} inferred
an oblate geometry  for the galaxy mass distribution. Also, recently obtained
ground-based images provide compelling evidence that NGC~205 indeed hosts an
embedded stellar disk (Saviane et al., in preparation).  A revised
classification from dE to a disk galaxy would provide a natural explanation for
many of the NGC~205 peculiarities, including the presence of gas and the
continuous star formation occurring in its central regions  during (at least)
the last $\sim 650$ Myr, as revealed by the present study.

\begin{acknowledgements}
M.B. and S.P acknowledge the financial support of INAF through the PRIN 2007
grant CRA 1.06.10.04. 
\end{acknowledgements}


\begin{thebibliography}{}
\bibitem[Bica et al.(1990)]{bica90} Bica, E., Alloin, D., \& 
Schmidt, A.~A.\ 1990, \aap, 228, 23 
\bibitem[Burstein et al.(1988)]{burstein88} 
Burstein, D., Bertola, F., Buson, et al.,\ 1988, \apj, 328, 440 (B88)
\bibitem[Butler \& Mart{\'{\i}}nez-Delgado(2005)]{butler05} 
Butler, D.~J., \& Mart{\'{\i}}nez-Delgado, D.\ 2005, \aj, 129, 2217 (BM05)
\bibitem[Buzzoni(2002)]{buzzoni02}
Buzzoni, A.\ 2002, \aj, 123, 1188
\bibitem[Cappellari et al.(1999)]{cappellari99} Cappellari, M., 
et al.\ 1999, \apjl, 515, L17 (C99)
\bibitem[Cepa \& Beckman(1988)]{cepa88} Cepa, J., \& Beckman, J.~E.\ 1988, 
\aap, 200, 21 (CP88)
\bibitem[Davidge(2003)]{davidge03} Davidge, T.~J.\ 2003, \apj, 597, 289 (D03)
\bibitem[Davidge(2005)]{davidge05} Davidge, T.~J.\ 2005, \aj, 130, 2087 
\bibitem[De Rijcke et al.(2006)]{der06} De Rijcke, S., et al.,\ 2006, 
\mnras, 369, 1321 
\bibitem[Galleti et al.(2006)]{silvia} Galleti, S., et al., 2006, 
\apj, 650, L107
\bibitem[Geha et al.(2006)]{geha06} Geha, M., Guhathakurta, 
P., Rich, R.~M., \& Cooper, M.~C.\ 2006, \aj, 131, 332 
\bibitem[Gerola \& Seiden(1978)]{gs78} Gerola, H., \& Seiden, P.~E.\ 1978, 
\apj, 223, 129 
\bibitem[Girardi et al.(2002)]{leo} Girardi, L., et al.\ 2002, \aap, 391, 195 
\bibitem[Harris(1996)]{harris} Harris, W.~E.\ 1996, \aj, 112, 
1487
\bibitem[Hodge(1973)]{hodge73} Hodge, P.~W.\ 1973, \apj, 182, 671 
\bibitem[Howley et al.(2008)]{howley} Howley, K.~M., et al.\ 2008, \apj, 683,
722 
\bibitem[Jones et al.(1996)]{jones} Jones, D.~H., et al.\ 1996, \apj, 466, 742 
\bibitem[Lee(1996)]{lee96} Lee, M.~G.\ 1996, \aj, 112, 1438 
\bibitem[Lisker et al.(2006)]{lisker_etal06} Lisker, T., Grebel, 
E.~K., \& Binggeli, B.\ 2006, \aj, 132, 497 
\bibitem[Lisker et al.(2007)]{lisker_etal07} Lisker, T., Grebel, 
E.~K., Binggeli, B., \& Glatt, K.\ 2007, \apj, 660, 1186 
\bibitem[Marleau et al.(2006)]{marleau06} Marleau, F.~R., et al.\ 
2006, \apj, 646, 929
\bibitem[Mateo(1998)]{mateo98} Mateo, M.~L.\ 1998, \araa, 36, 435 
\bibitem[McConnachie et al.(2004)]{mccon04} McConnachie, A.~W., 
et al.\ 2004, \mnras, 351, L94
\bibitem[McConnachie et al.(2005)]{mccon05} McConnachie, A.~W., 
et al.\ 2005, \mnras, 356, 979 
\bibitem[Mould et al.(1984)]{mould84} Mould, J., Kristian, J., 
\& Da Costa, G.~S.\ 1984, \apj, 278, 575 
\bibitem[Peletier (1993)]{peletier} Peletier, R.F., 1993, \aap, 271, 51
\bibitem[Perina et al.(2009)]{sibilla} Perina, S., et al.\ 2009, \aap, 494, 933 
\bibitem[Richer et al.(1984)]{richer84} Richer, H.~B., Crabtree, 
D.~R., \& Pritchet, C.~J.\ 1984, \apj, 287, 138 
\bibitem[Rossa et al.(2006)]{rossa} Rossa, J., van der Marel, R.P., B\"oker, T.,
et al., 2006, \pasp, 109, 907
\bibitem[Saviane et al.(2004)]{sfr} Saviane, I., Hibbard, J. E.,  Rich,
R. M. 2004, AJ, 127, 660 (SHR04)	
\bibitem[Saviane et al.(2000)]{ivo00} Saviane, I., Rosenberg, A., Piotto, G., \&
Aparicio, A.\ 2000, \aap, 355, 966 
\bibitem[Schlegel et al.(1998)]{sfd} Schlegel, D.~J., 
Finkbeiner, D.~P., \& Davis, M.\ 1998, \apj, 500, 525
\bibitem[Sharina et al.(2006)]{sharina06} Sharina, M.~E., 
Afanasiev, V.~L., \& Puzia, T.~H.\ 2006, \mnras, 372, 1259 
\bibitem[Shore(1983)]{shore83} Shore, S.~N.\ 1983, \apj, 265, 202 
\bibitem[Sirianni et al.(2005)]{sirianni} Sirianni, M., Jee, M.J., Bemitez, N.,
et al., 2005, \pasp, 117, 1049 
\bibitem[Valle et al.(2005)]{valle05} Valle, G., Cignoni, M., \& 
Shore, S.~N.\ 2005, \aap, 440, 473 
\bibitem[Valluri et al.(2005)]{acs} Valluri, M., Ferrarese, 
L., Merritt, D., \& Joseph, C.~L.\ 2005, \apj, 628, 137
\bibitem[Welch et al.(1998)]{welch98} Welch, G.~A., Sage, 
L.~J., \& Mitchell, G.~F.\ 1998, \apj, 499, 209 
\bibitem[Wilcots et al.(1990)]{iue} Wilcots, E.~M., Hodge, 
P., Eskridge, et al,\ 1990, \apj, 364, 87 
\end{thebibliography}
\end{document}